\documentclass[letterpaper]{article}
\usepackage[utf8]{inputenc}
\usepackage{geometry}
\usepackage{setspace}
\usepackage[style = chem-acs]{biblatex}
\usepackage[T1]{fontenc}
\usepackage{xr-hyper}
\usepackage{graphicx}
\usepackage{bm}
\usepackage{amsmath}
\usepackage{float}
\usepackage{caption}
\usepackage{subcaption}
\usepackage{multirow}
\usepackage{amsfonts}
\usepackage{textcomp, gensymb}
\usepackage{algpseudocode}
\usepackage{algorithm}
\usepackage[normalem]{ulem} 
\usepackage{hyperref}
\hypersetup{
    colorlinks=true,
    linkcolor=blue,
    filecolor=magenta,
    urlcolor=cyan,
    pdfpagemode=FullScreen,
    }
\usepackage{makecell}
\usepackage{xcolor}
\usepackage{appendix}
\usepackage{url}
\usepackage{amsthm}
\usepackage{comment}
\usepackage{authblk}
\usepackage{ulem}

\newcommand{\angstrom}{\mbox{\normalfont\AA}}

\newcommand{\Eqn}[1]{Eq~(\ref{#1})}
\newcommand{\eqn}[1]{eq~(\ref{#1})}
\newcommand{\eqns}[1]{eqs~(\ref{#1})}
\newcommand{\eqnx}[1]{(\ref{#1})}

\newcommand{\sect}[1]{Section~\ref{#1}}
\newcommand{\sects}[1]{Sections~\ref{#1}}
\newcommand{\sectx}[1]{\ref{#1}}

\newcommand{\fig}[1]{Figure~\ref{#1}}

\theoremstyle{definition}

\title{Statistical mechanics of a 2D material in a gas reservoir}
\author[1,2]{Moon-ki~Choi}
\author[3]{Ellad~B.~Tadmor*}
\affil[1]{\small Department of Mechanical Science and Engineering, University of Illinois Urbana-Champaign, 1206 W. Green St, Urbana, IL, 61801, USA}
\affil[2]{\small Materials Research Laboratory, University of Illinois Urbana-Champaign, 104 South Goodwin Ave. MC-230, Urbana, IL, 61801, USA}
\affil[3]{\small Department of Aerospace Engineering and Mechanics, University of Minnesota, 110 Union St SE, Minneapolis, MN, 55455, USA}
\date{*Email: tadmor@umn.edu}
\graphicspath{{figures/}{./}}

\date{}
\begin{document}
\maketitle
\begin{abstract}
We derive and validate a partition function for low-dimensional systems interacting with a heat bath, addressing the general issue of thermodynamic modeling of nanoscale systems. In contrast to bulk systems in the canonical (NVT) ensemble where the partition function is solely determined by the Hamiltonian of the system and the temperature of the heat bath, our formulation demonstrates that accounting for the interactions with the heat bath is essential for describing the statistical mechanics of low-dimensional materials. To validate our theoretical findings, we develop a molecular dynamics (MD) algorithm for directly modeling the heat bath as a gas reservoir. We first validate our approach using a 1D harmonic oscillator, calculating its length distribution through explicit numerical integration and confirming these results with MD simulations. We then extend our method to investigate the out-of-plane fluctuations of a 2D graphene monolayer immersed in a gas at finite temperature and pressure. Comparisons with conventional NVT ensemble simulations controlled by a thermostat reveal that environmental interactions significantly influence the properties of the 2D material system.
\end{abstract}

\section{Introduction}
\label{sec:intro}
The canonical ensemble represents the equilibrium thermodynamic behavior of a system in contact with a heat bath. Its derivation relies on the assumption that the system is macroscopic and yet significantly smaller in size compared to the heat bath. A further assumption is that the system and heat bath are in \emph{weak interaction}, meaning that the interaction energy between them can be neglected when calculating statistical mechanics phase averages of the system \cite{tadmor2011,tuckerman2010,reif2009}. Gibbs explained it in this way \cite{gibbs1902}:
\begin{quote}
The most simple test of the equality of temperature of two bodies is that they remain in equilibrium when brought into thermal contact. Direct thermal contact implies molecular forces acting between the bodies. Now the test will fail unless the energy of these forces can be neglected in comparison with the other energies of the bodies. Thus, in the case of energetic chemical action between the bodies, or when the number of particles affected by the forces acting between the bodies is not negligible in comparison with the whole number of particles (as when the bodies have the form of exceedingly thin sheets), the contact of bodies of the same temperature may produce considerable thermal disturbance, and thus fail to afford a reliable criterion of the equality of temperature.
\end{quote}
This immediately raises the question as to whether the canonical ensemble, and NVT thermostats based on it that are used in molecular dynamics (MD) simulations to maintain temperature, are applicable to two-dimensional (2D) materials, which are the most extreme case of ``exceedingly thin sheets.'' In a 2D material, the number of particles affected by the heat bath is equal to the total number of particles. Clearly, the assumption of weak interaction can fail in this case if the interaction energy significantly impacts phase averages of the system. This topic is part of the larger question of the thermodynamics of strongly coupled systems \cite{seifert2016,jarzynski2017,Talkner2020-zx,du2025}.

To explore this question, we revisit the canonical ensemble explicitly accounting for the possibility of strong interactions with a heat bath. The effect of these interactions is captured through an additional term in the canonical ensemble distribution function. In addition, we develop a \textit{gas reservoir thermostat} for MD simulations that maintains the temperature of a system by interfacing it with a heat bath through an intermediary ideal gas. The algorithm allows for the inward and outward flow of gas atoms at simulation cell boundaries, effectively modeling contact with an infinite heat bath within a finite simulation domain. The thermodynamic equilibrium properties of the simulated gas reservoir, including velocity, number of atoms, and temperature are designed to match ideal gas properties.

We study the new partition function and thermostat through three representative examples. First, we consider a one-dimensional (1D) harmonic oscillator in contact with a heat bath. An analytical expression is obtained for the distribution function, and the bond length distributions for different temperatures and pressures of the gas reservoir are computed via numerical integration. These computed distributions are subsequently validated via MD simulation of the harmonic oscillator with the gas reservoir thermostat.

Second, we investigate the thermal equilibration of a gold (Au) nanoparticle, showing that the relaxation dynamics depend on the reservoir pressure. After equilibration, we find increasing deviations from canonical predictions for the effective particle radius as the pressure increases.

Third, we explore the properties of a 2D material in contact with a heat bath. Specifically, we perform MD simulations of a graphene monolayer at constant temperature and pressure and compute its out-of-plane fluctuation at equilibrium. This property is assessed in two ways: 1) by applying the conventional canonical ensemble with an NVT thermostat, and 2) by employing the new gas reservoir thermostat. The observed differences between these two conditions highlight the influence of non-negligible  heat bath interactions on the properties of a 2D material.

The paper is organized as follows: \sect{sec:stat_mec_2D} derives the probability density function for a system interacting strongly with a heat bath. \sect{sec:gas_res} describes the proposed MD algorithm for a gas reservoir and discusses its properties. \sect{sec:num_harmonic} presents the application of the gas reservoir simulation to a 1D harmonic oscillator. \sect{sec:np_eq} examines the equilibrium of a Au nanoparticle in the gas reservoir, and \sect{sec:gra_prop} analyzes the properties of graphene in the gas reservoir. The paper concludes in \sect{sec:conclusion} with a summary and suggestions for future work.

\section{Canonical ensemble accounting for interactions with a heat bath}
\label{sec:stat_mec_2D}
Based on the assumption of weak interaction, the probability density function (PDF) of a system in contact with a heat bath is \cite{tadmor2011,tuckerman2010}
\begin{equation}\label{eqn:prob}
    f(\bm{q},\bm{p};T) = \frac{1}{Z}\exp{\bigl[-\beta\mathcal{H}(\bm{q},\bm{p})\bigr]},
\end{equation}
where $Z$ is the partition function,
\begin{equation}
    Z = \int_{\Gamma_{\bm{q},\bm{p}}} \exp{\bigl[-\beta\mathcal{H}(\bm{q},\bm{p})\bigr]}d\bm{q}d\bm{p}.
\end{equation}
Here, $\bm{q}$ and $\bm{p}$ are the positions and linear momenta of the atoms in the system, $\Gamma_{\bm{q},\bm{p}}$ is the accessible phase space of $\bm{q}$ and $\bm{p}$, $\mathcal{H}$ is the Hamiltonian of the system, and $\beta = 1/(k_{\rm B}T)$ where $k_B$ is the Boltzmann constant, and $T$ is the temperature of the heat bath. Note that the PDF in \eqn{eqn:prob} depends only on the Hamiltonian of the system. We refer to the canonical ensemble associated with \eqn{eqn:prob} as a ``conventional'' canonical ensemble to distinguish it from the derivation below that does not make the weak interaction assumption.

We now consider the case where the PDF in \eqn{eqn:prob} is applied to a composite system comprised of a system S contained within a gas reservoir R, and both S and R are in contact with a much larger heat bath B. Weak interactions are assumed between B and the combined system S and R, while the interactions between S and R are not ignored.\footnote{This approach resembles the derivation of an isothermal-isobaric (NPT) partition function, where two systems are in contact with a heat bath as explained in \cite{tuckerman2010}. However, our approach incorporates the interactions between S and R, which are neglected in the derivation of the NPT partition function. Additionally, here we assume that S is significantly smaller than R, so that volume changes of S do not influence the volume of R. This is in contrast to the NPT partition function derivation, where the volume change of S is non-negligible.}
The number of atoms in S ($N_{\rm S}$) is much smaller than that of R ($N_{\rm R}$), and both $N_{\rm S}$ and $N_{\rm R}$ are much smaller than the number of atoms in the heat bath ($N_{\rm B}$). i.e., $N_{\rm S} \ll N_{\rm R} \ll N_{\rm B}$.
Given this condition, the PDF of S and R using \eqn{eqn:prob} is
\begin{equation}\label{eqn:prob_SR}
    f_{\rm SR}(\bm{q}^{\rm S},\bm{p}^{\rm S},\bm{q}^{\rm R},\bm{p}^{\rm R};T) =
    \frac{1}{Z_{\rm SR}}\exp{\Bigl[-\beta\bigl(\mathcal{H}_{\rm S}(\bm{q}^{\rm S},\bm{p}^{\rm S})+\mathcal{V}_{\rm SR}(\bm{q}^{\rm S},\bm{q}^{\rm R})+\mathcal{H}_{\rm R}(\bm{q}^{\rm R},\bm{p}^{\rm R})\bigr)\Bigr]},
\end{equation}
where the partition function ($Z_{\rm SR}$) is given by
\begin{equation}\label{eqn:Z_SR}
    Z_{\rm SR} = \int_{\Gamma_{\bm{q}^{\rm S},\bm{p}^{\rm S},\bm{q}^{\rm R},\bm{p}^{\rm R}}} \exp{\Bigl[-\beta\bigl(\mathcal{H}_{\rm S}+\mathcal{V}_{\rm SR}+\mathcal{H}_{\rm R}\bigr)\Bigr]}\
    d\bm{q}^{\rm S}d\bm{p}^{\rm S}d\bm{q}^{\rm R}d\bm{p}^{\rm R}.
\end{equation}
In \eqns{eqn:prob_SR} and \eqnx{eqn:Z_SR}, $\bm{q}^{\rm S}$ and $\bm{q}^{\rm R}$ are the positions, and $\bm{p}^{\rm S}$ and $\bm{p}^{\rm R}$ are the momenta, of the atoms in S and R. $\mathcal{H}_{\rm S} (\bm{q}^{\rm S},\bm{p}^{\rm S})$ and $\mathcal{H}_{\rm R} (\bm{q}^{\rm R},\bm{p}^{\rm R})$ are the Hamiltonians of S and R on their own, and $\mathcal{V}_{\rm SR} (\bm{q}^{\rm S},\bm{q}^{\rm R})$ is the interaction energy between them. To keep the notation concise, the arguments of $\mathcal{H}_{\rm S}$, $\mathcal{H}_{\rm R}$, and $\mathcal{V}_{\rm SR}$ are omitted in \eqn{eqn:Z_SR} and in following expressions.

By integrating \eqn{eqn:prob_SR} over $\bm{q}^{\rm R}$ and $\bm{p}^{\rm R}
$, the PDF of S follows as
\begin{align}\label{eqn:prob_2D}
    f_{\rm S}(\bm{q}^{\rm S},\bm{p}^{\rm S};T) &=
    \frac{1}{Z_{\rm SR}}\exp{(-\beta\mathcal{H}_{\rm S})}
    \int_{\Gamma_{\bm{q}^{\rm R},\bm{p}^{\rm R}}} \exp{\bigl[-\beta(\mathcal{V}_{\rm SR}+\mathcal{V}_{\rm R}+\mathcal{K}_{\rm R})\bigr]}
    d\bm{q}^{\rm R}d\bm{p}^{\rm R} \\ \nonumber
    & = \frac{1}{Z_{\rm SR}}\exp{(-\beta\mathcal{H}_{\rm S})} C(T)
    \int_{\Gamma_{\bm{q}^{\rm R}}} \exp{\bigl[-\beta( \mathcal{V}_{\rm SR}+\mathcal{V}_{\rm R})\bigr]}  d\bm{q}^{\rm R}
    \\  \nonumber
    & = \frac{1}{Z_{\rm SR}}\exp{(-\beta\mathcal{H}_{\rm S})}
    C(T)
    \Phi(\bm{q}^{\rm S};T),
\end{align}
where
\begin{align}\label{eqn:prob_2D_2}
     C(T) &= \int_{\Gamma_{\bm{p}^{\rm R}}} \exp{(-\beta \mathcal{K}_{\rm R})} d\bm{p}^{\rm R},
\\
     \Phi(\bm{q}^{\rm S};T) &=
    \int_{\Gamma_{\bm{q}^{\rm R}}}\exp{\bigl[-\beta(\mathcal{V}_{\rm SR}+\mathcal{V}_{\rm R})\bigr]}d\bm{q}^{\rm R}.
\end{align}
Here, $\mathcal{K}_{\rm R}$ and $\mathcal{V}_{\rm R}$ are the kinetic and potential energies of R, respectively. Comparing the PDF in \eqn{eqn:prob_2D} with the conventional expression in \eqn{eqn:prob}, we see that $f_{\rm S}$ includes two additional terms: $\Phi\left(\bm{q}^{\rm S};T\right)$ and $C(T)$. $C(T)$ is a constant at a given $T$, and therefore just scales the distribution function without changing its form. In contrast, $\Phi(\bm{q}^{\rm S};T)$ has a direct effect on the form of $f_{\rm S}$ through its dependence on $\mathcal{V}_{\rm SR}$. In the Supporting Information (SI), we study the equipartition theorem for $f_{\rm S}$, considering both the positions and momenta of S, and compare the results with those obtained from the conventional NVT ensemble (see Section S.1).

Obtaining a closed-form expression for $f_{\rm S}(\bm{q}^{\rm S},\bm{p}^{\rm S};T)$ is not possible for general nonlinear interaction potentials, however in \sect{sec:num_harmonic}, we validate \eqn{eqn:prob_2D} for a simple 1D harmonic oscillator embedded in a gas reservoir. Then in \sects{sec:np_eq} and \sectx{sec:gra_prop}, we numerically explore how \eqn{eqn:prob_2D} influences the properties of a Au nanoparticle and a graphene monolayer in a gas reservoir as compared with a conventional NVT ensemble governed by \eqn{eqn:prob}.

\section{Gas reservoir simulation}
\label{sec:gas_res}
\subsection{Method}
\label{sec:gas_res_method}
In principle, MD simulations of a system S in contact with a heat bath $B$ can be performed by embedding the system within a surrounding gas in which the gas atoms interact with the system using a suitable interatomic potential.\footnote{Unless there is interest in modeling reactions between the gas and system, for example surface oxidation, an inert gas can be chosen that interacts with the surface via van der Waals forces, which can be modeled using a Lennard--Jones potential.} However the number of gas atoms in the heat bath would have to be very large relative to the system ($N_{\rm B}\gg N_{\rm S}$) in order for the heat bath to maintain a constant temperature irrespective of S making it computationally impractical. Instead, here we develop an algorithm for simulating a system in contact with an infinite heat bath via an intermediate gas reservoir along the lines described in \sect{sec:stat_mec_2D}. The approach utilizes a surface generation reservoir technique originally designed for direct simulation Monte Carlo (DSMC) by Tysanner and Garcia \cite{tysanner2005}. In this method, gas atoms are introduced into a simulation box through a boundary surface with an average inward flux governed by equilibrium gas dynamics. The expected mean number of atoms entering the simulation box within a time interval $\Delta t$ is given by
\begin{equation}\label{eqn:numgas}
    \left<N_{\rightarrow}\right> = n_{\rm R}A_{\rm R} \sqrt{\frac{k_{\rm B}T_{\rm R}}{2\pi m}}\Delta t,
\end{equation}
where $n_{\rm R}$ is the number density of the gas reservoir, $A_{\rm R}$ is the area of the boundary surface, $T_{\rm R}$ is the reservoir temperature, and $m$ is the mass of a gas atom.
\Eqn{eqn:numgas} is rewritten using the ideal gas law as
\begin{equation}\label{eqn:numgasp}
    \left<N_{\rightarrow}\right> = \frac{p_{\rm R}}{k_{\rm B}T_{\rm R}}A_{\rm R} \sqrt{\frac{k_{\rm B}T_{\rm R}}{2\pi m}}\Delta t,
\end{equation}
where $p_{\rm R}$ is the pressure of the gas reservoir.
Simply injecting atoms into a simulation at each time step $\Delta t$ according to \eqn{eqn:numgasp} leads to an anomalous velocity profile in the gas reservoir \cite{tysanner2005}. Instead, the actual number of gas atoms introduced is drawn from a Poisson distribution with  mean $\left<N_{\rightarrow}\right>$:
\begin{equation}\label{eqn:poi}
    P(k; \left<N_{\rightarrow}\right>) = \frac{\left<N_{\rightarrow}\right>^k e^{-\left<N_{\rightarrow}\right>}}{k!},
\end{equation}
where $k$ is the actual number of particles introduced during the time interval $\Delta t$.

\begin{figure}\centering
    \includegraphics[width=1.0\textwidth]{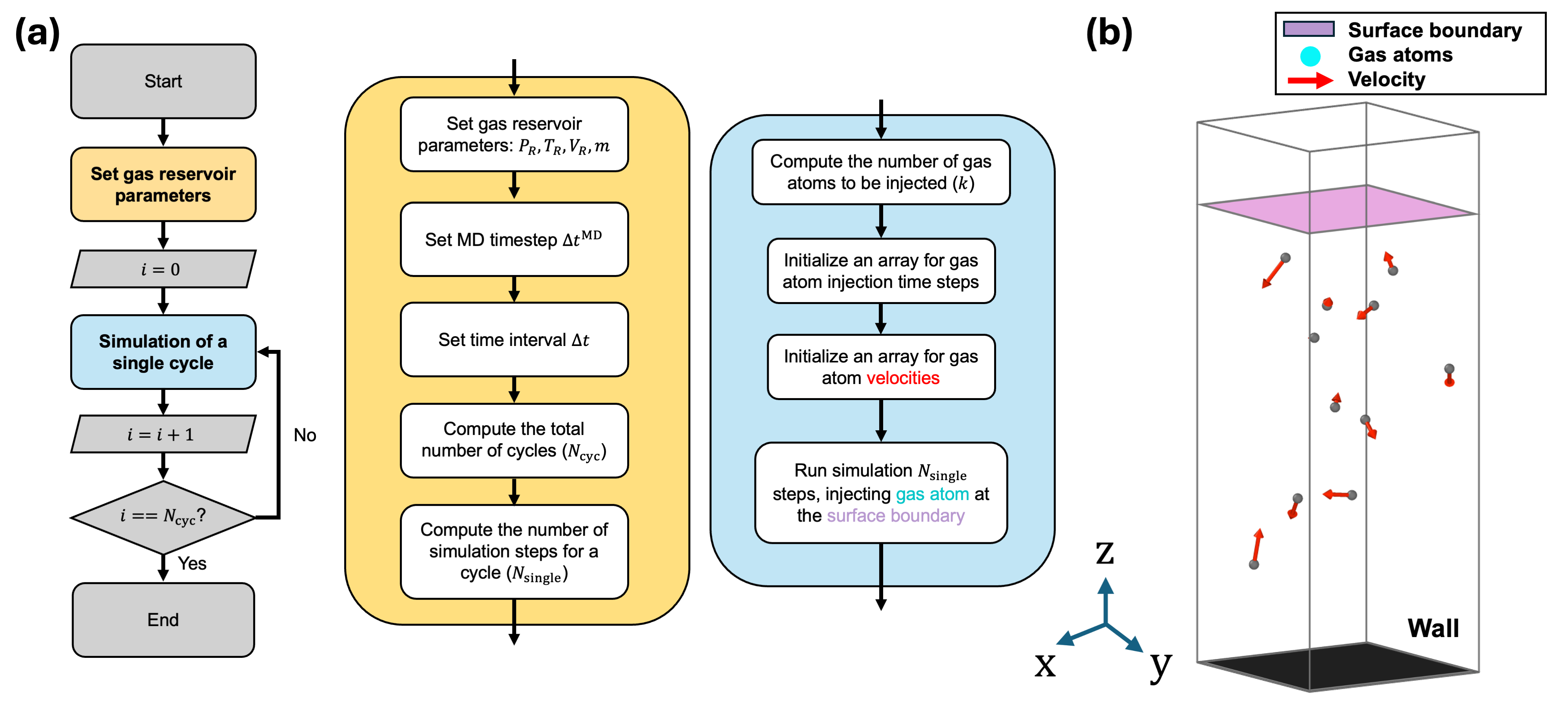}
    \caption{(a) Flow chart of the gas reservoir thermostat algorithm. (b) Snapshot of an MD simulation with the gas reservoir thermostat.}
    \label{fig:flowchart}
\end{figure}

\fig{fig:flowchart}a illustrates the flowchart for the gas reservoir thermostat algorithm, highlighting two primary processes: 1) setting of gas reservoir parameters (orange box), and 2) simulation of a single cycle (blue box). In the ``set gas reservoir parameters'' stage, the gas reservoir simulation conditions, $p_{\rm R}, T_{\rm R}$, $\Delta t$, $V_{\rm R}$, and $m$. are specified. The time interval for a single gas particle injection cycle ($\Delta t$) is chosen based on \eqn{eqn:numgasp} to ensure a sufficiently large value for $\left<N_{\rightarrow}\right>$ (i.e., $\left<N_{\rightarrow}\right> \gg 1$) at the specified reservoir parameters.  For a given total simulation time, $t^{\rm  total}$, the number of simulation cycles ($N_{\rm cyc}$) is calculated as
\begin{equation}
N_{\rm cyc}=\text{round}\left(\frac{t^{\rm total}}{\Delta t}\right) .
\end{equation}
Also, the number of simulation steps within a single cycle is given by
\begin{equation}
N_{\rm single}=\text{round}\left(\frac{\Delta t}{ \Delta t^{\rm MD}}\right),
\end{equation}
where $\Delta t^{\rm MD}$ is the time step used in the MD simulation. After computing $N_{\rm cyc}$ and $N_{\rm single}$, the cycle index $i$ is initialized to 0. The single cycle process is repeated until  $i$ equals $N_{\rm cyc}$.

In the ``simulation of a single cycle'' process, gas atoms are injected at the surface over the time $\Delta t$.  The total number of atoms entering the reservoir surface within the time interval $\Delta t$, $N_{\Delta t}=k$, is computed using \eqn{eqn:poi}. An array with $k$ elements is then initialized to store the specific insertion time for each atom. These $k$ timing slots are randomly selected from $\{0,1,\ldots,N_{\rm single}\}$. During the simulation, each gas atom is injected on the surface at its predetermined time, at a location randomly assigned from a uniform distribution. The velocity component perpendicular to the surface (denoted $v_\perp$), pointing toward the interior of the simulation box, is drawn from a biased Maxwell-Boltzmann distribution, expressed as
\begin{equation}\label{eqn:biased_MB}
f_\perp(v_\perp) = \frac{m}{k_{\rm B}T_{\rm R}}v_{\perp}
\exp{\biggl(\frac{-mv_{\perp}^2}{2k_{\rm B}T_{\rm R}}\biggr)}.
\end{equation}
The velocity components in the in-plane direction ($v_{x}$ and $v_{y}$) are drawn from a Maxwell-Boltzmann distribution:
\begin{equation}\label{eqn:MB}
f_\parallel(v_\parallel) = \sqrt{\frac{m}{2\pi k_B T_{\rm R}}}
\exp{\biggl(\frac{-mv_{\parallel}^2}{2k_{\rm B}T_{\rm R}}\biggr)},
\end{equation}
where $v_{\parallel}$ represents either $v_{x}$ or $v_{y}$. After sampling the magnitudes of $v_x$ and $v_y$ from \eqn{eqn:MB},
the sign of each component is assigned randomly with equal probability for $+$ and $-$.

The biased Maxwell--Boltzmann distribution in \eqn{eqn:biased_MB} arises directly from kinetic theory. In an equilibrium gas, the flux of particles crossing a planar surface is proportional to the perpendicular speed $v_\perp$ multiplied by the Maxwell--Boltzmann velocity distribution; particles with larger $v_\perp$ are therefore more likely to reach and cross the boundary within a given time interval. Normalizing this flux-weighted distribution over $v_\perp>0$ results in \eqn{eqn:biased_MB}. In contrast, the in-plane components ($v_x$ and $v_y$) are not conditioned on the surface crossing and therefore follow the Maxwell--Boltzmann distribution in \eqn{eqn:MB}. When combined, these sampling rules ensure that the gas atoms inside follow the Maxwell--Boltzmann velocity distribution of the reservoir.

Gas atoms leaving the simulation box through the reservoir boundary with outward-pointing velocity components are removed from the system, ensuring an open boundary consistent with an infinite heat bath.

\fig{fig:flowchart}b shows a snapshot of a gas reservoir simulation following the algorithm in \fig{fig:flowchart}a.  Gas atoms (cyan points) are injected from the boundary surface (purple plane), with their velocity shown as a red arrow. Details of this MD simulation are discussed in \sect{sec:gas_res_num}.
The gas reservoir MD simulation is performed using a Python script interfaced with LAMMPS \cite{thompson2022}. The Python code for running the gas reservoir simulation is available in \cite{gascode}.

\subsection{Properties of the gas reservoir}
\label{sec:gas_res_num}
To test that the formulation in \sect{sec:gas_res_method} achieves the expected state of thermodynamic equilibrium, we perform MD with a simulation box of dimensions 300 $\times$ 300 $\times$ 600$~\angstrom^3$. The gas reservoir surface is located at 500~$\angstrom$, with periodic boundary conditions (PBC) applied in $x$ and $y$ directions, and a reflective wall located at $z=0$. This results in an actual gas reservoir size of 300 $\times$ 300 $\times$ 500$~\angstrom^3$.\footnote{To prevent gas atoms that exit the reservoir boundary surface from reappearing on the opposite side due to PBCs of the simulation, we employ the \texttt{fix evaporate} command in LAMMPS. This command effectively removes gas atoms that move beyond the reservoir boundary (i.e., the purple surface in \fig{fig:flowchart}b).} The gas is taken to be argon (Ar) with Lennard--Jones (LJ) interactions between the atoms given by
\begin{equation}\label{eqn:LJ}
V(r) = 4\epsilon\biggl[\Bigl(\frac{\sigma}{r}\Bigr)^{12}-\Bigl(\frac{\sigma}{r}\Bigr)^{6}\biggr],
\end{equation}
where $r$ is the distance between atoms, $\epsilon$ is the LJ energy parameter, and $\sigma$ is the LJ distance parameter. We set $\epsilon=1.03 \times 10^{-2} $ eV and $\sigma =3.405\;\angstrom$ \cite{yarnell1973}. The cutoff distance of the LJ potential is taken to be 12~$\angstrom$, and the time step of the simulation is 10 fs. We choose $\Delta t=10$~ns, resulting in $\langle N_{\rightarrow}\rangle=97.54$ using \eqn{eqn:numgasp} at $T_{\rm R}=300$~K and $p_{\rm R}=1$~atm.

\begin{figure}\centering
    \includegraphics[width=0.6\textwidth]{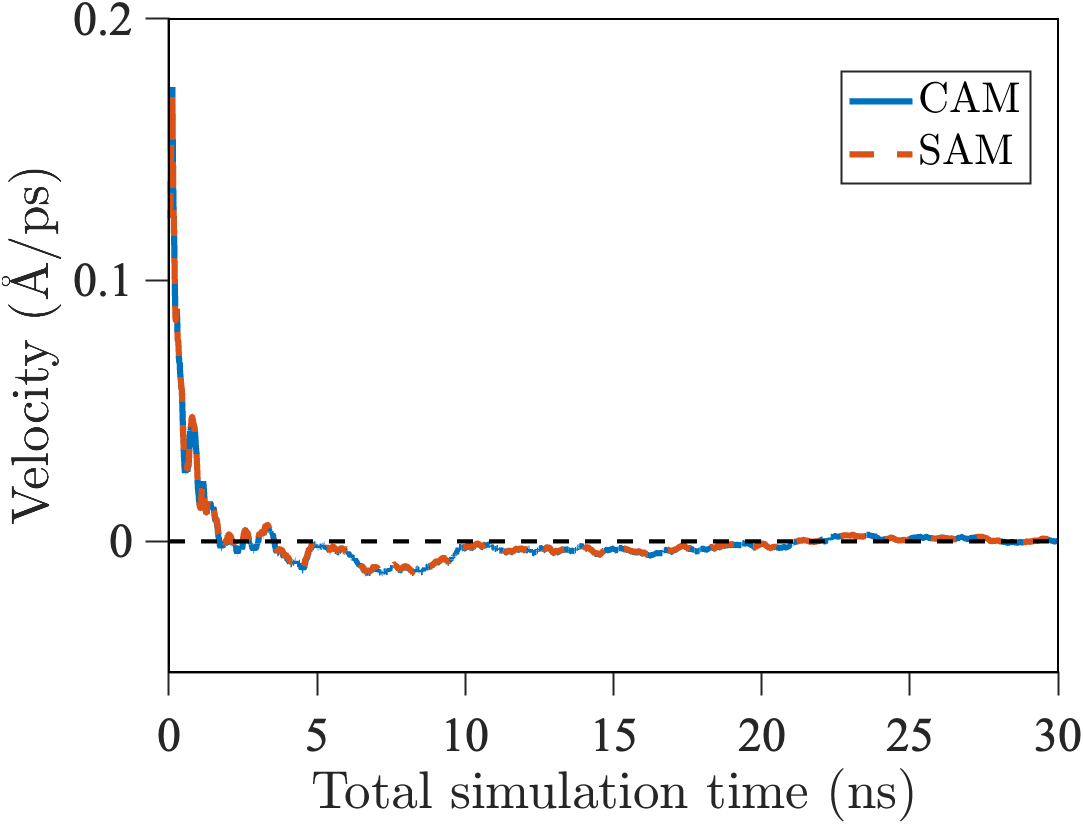}
    \caption{CAM (blue line) and SAM (red dotted line) of the gas reservoir.}
    \label{fig:SAM_CAM}
\end{figure}

Tysanner and Garcia \cite{tysanner2004,tysanner2005} evaluate the convergence of the gas reservoir in DSMC simulation by computing a cumulative average measurement (CAM) and a sample averaging measurement (SAM).
CAM computes the mean velocity as the ratio of the cumulative momentum to the cumulative mass in each cell, whereas SAM averages the instantaneous cell velocity at each sampling time. Because SAM is sensitive to correlations between fluctuations in density and momentum, it can exhibit a bias under non-equilibrium conditions, while CAM remains unbiased. The formal definitions of CAM and SAM are provided in Eqs.~(1)--(4) of \cite{tysanner2005} along with computation algorithms.
Both the SAM and CAM values should converge to zero when the gas reservoir is in equilibrium, indicating zero velocity and momentum of the gas reservoir system at equilibrium.

\fig{fig:SAM_CAM} shows the CAM and SAM of the velocity in the $z$ direction for the gas atoms as a function of the simulation time.
We see that both SAM and CAM converge to zero after approximately 20~ns, indicating that the gas reservoir has reached thermodynamic equilibrium.
We also note that the disparity between CAM and SAM is close to zero, indicating that the simulated gas reservoir is in equilibrium in terms of both velocity and momentum \cite{tysanner2004,tysanner2005}.

\begin{figure}
\begin{center}
\subfloat[$N_{\rm sim}/N_{\rm R}$]{\includegraphics[trim = 0mm 0mm 0mm 0mm, clip=true,width=0.49\textwidth]{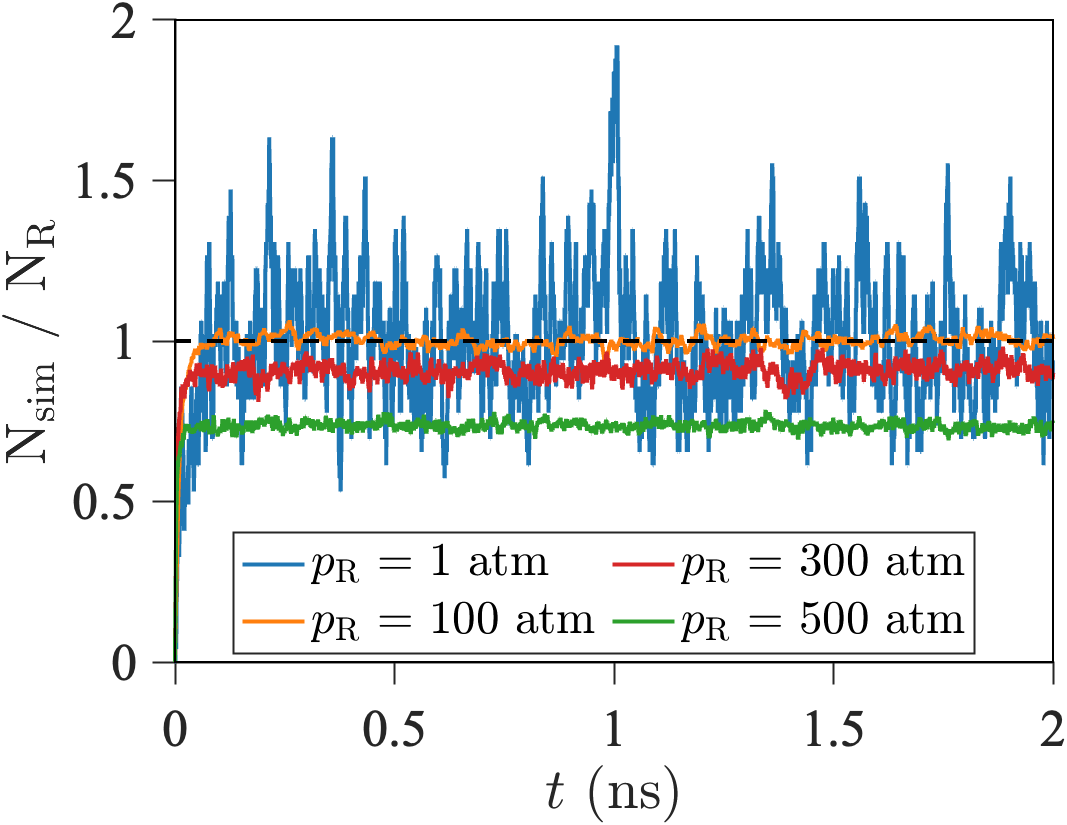}\label{fig:natom_t}}\hfill
\subfloat[$T_{\rm sim}/T_{\rm R}$]{\includegraphics[trim = 0mm 0mm 0mm 0mm, clip=true,width=0.49\textwidth]{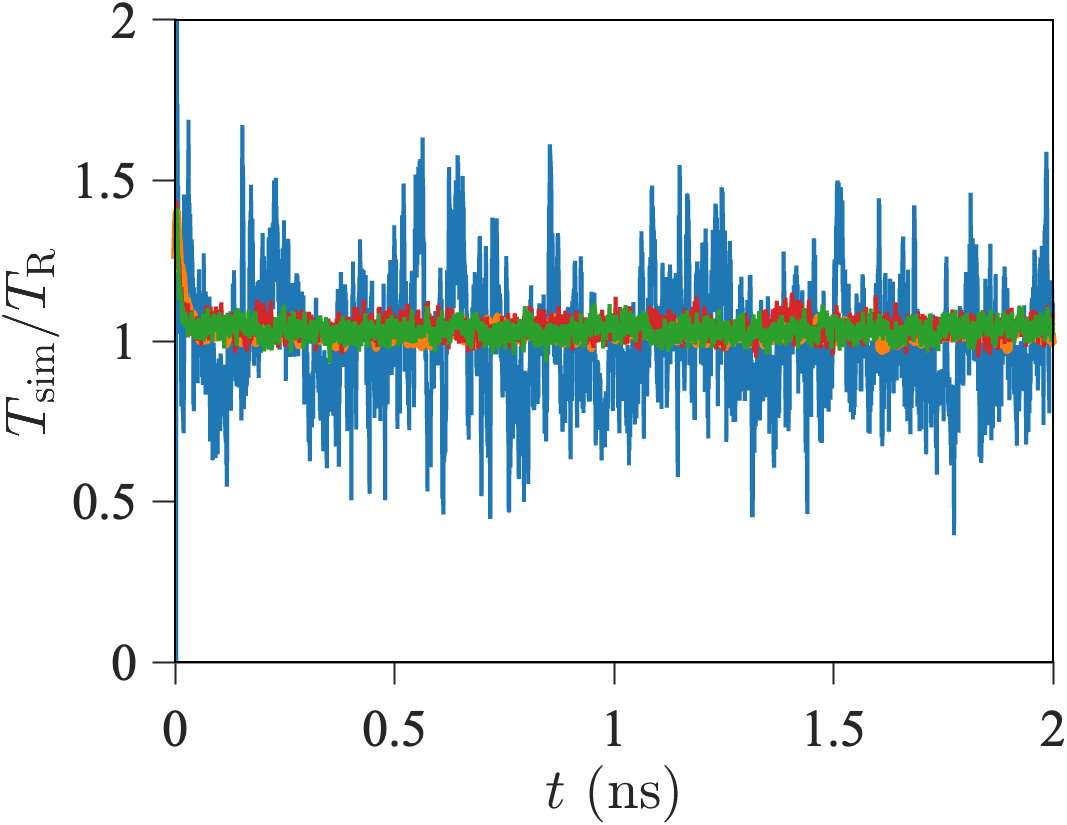}\label{fig:temp_t}}\hfill
\caption{$N_{\rm sim}/N_{\rm R}$ and $T_{\rm sim}/T_{\rm R}$ of the gas reservoir simulation over a simulation of 2 ns.}
\label{fig:res_prop}
\end{center}
\end{figure}

To further examine the equilibrium state of the gas reservoir MD simulation, we calculate both the number of gas atoms ($N_{\rm sim}$) and the simulation temperature ($T_{\rm sim}$). These are compared to the initial reservoir parameters; the atom count is determined from the ideal gas law:
\begin{equation}
    N_{\rm R} = \frac{p_{\rm R}V_{\rm R}}{k_{\rm B}T_{\rm R}}.
\end{equation}
This comparison is performed over a range of $p_{\rm R}$ values for the set value of  $T_{\rm R}$ to modulate the gas number density in the reservoir.
\fig{fig:res_prop} shows the computed ratios of $N_{\rm sim}/N_{\rm R}$ and $T_{\rm sim}/T_{\rm R}$ for the gas reservoir under various $p_{\rm R}$ values over 2 ns.  As shown in  \fig{fig:natom_t}, gas atoms rapidly fill the reservoir, with $N_{\rm sim}$ /$N_{\rm R}$ stabilizing within 0.2 ns for the tested pressures (1, 100, 300, and 500 atm). At $p_{\rm R}=1$~atm, the large fluctuations in $N_{\rm sim}$/$N_{\rm R}$ are due to the small number of gas atoms present.
At higher pressures the magnitude of the fluctuations reduce, and the average $N_{\rm sim}$/$N_{\rm R}$ drops below 1 (i.e., $N_{\rm sim}<N_{\rm R}$), decreasing with increasing pressure.
This decrease is attributed to the increased effective volume occupied by each gas atom. In particular, the effective interatomic spacing $(V_{\rm R}/N_{\rm R})^{1/3}$ for Ar is  34.45, 7.42, 5.14, and 4.34~\AA\ for pressures of 1, 100, 300, and 500~atm, respectively.
At pressures of 100~atm and above, these effective distances fall below the cutoff distance (12 $\angstrom$), resulting in a higher frequency of gas--gas collisions. Consequently, the increased collisions drive $N_{\rm sim}$/$N_{\rm R}<1$. Because the simulation is based on the ideal gas law that inherently neglects gas--gas interactions, the deviation observed in $N_{\rm sim}$/$N_{\rm R}$ at high pressures indicates a significant departure from the ideal gas behavior in the simulated reservoir.\footnote{Note that the relatively large MD time step used in the simulations means that the details of gas--gas interactions may not be accurately captured, affecting the accuracy of the reported values, but not the overall conclusions.}

Despite the observed non-ideal behavior, \fig{fig:temp_t} demonstrates that the average of $T_{\rm sim}/T_{\rm R}$ is close to 1, indicating that the expectation of $T_{\rm sim}$ remains constant and close to the set temperature ($T_{\rm R}=300$~K) for all $p_{\rm R}$ conditions.
At low $p_{\rm R}$, $T_{\rm sim}/T_{\rm R}$ shows a high fluctuation because $N_R$ is lower than other pressure cases.
We observe that $T_{\rm sim}/T_{\rm R}$ remains close to 1 at higher $p_{\rm R}$, despite the increased gas-gas interactions due to the reduced effective distance between gas atoms. This suggests that these interactions are primarily collisions rather than stable bond formations, conserving the kinetic energy even at high $p_{\rm R}$ conditions.

\section{Application to a 1D harmonic oscillator}\label{sec:num_harmonic}
We apply the gas reservoir thermostat formulation of \sect{sec:gas_res_method} to validate the PDF $f_{\rm S}$ (\eqn{eqn:prob_2D}) for the simple case of a 1D harmonic oscillator consisting of two atoms connected by a linear spring within a gas reservoir. The Hamiltonian is given by
\begin{equation}\label{eqn:H_har}
    \mathcal{H}_{\rm S} = \mathcal{V}_{\rm S} + \mathcal{K}_{\rm S} =a \vert\bm{q}^{\rm S}_1-\bm{q}^{\rm S}_2\vert^2 + \sum_{i=1}^2\frac{1}{2}m_h\vert\bm{p}^{\rm S}_i\vert^2,
\end{equation}
where $a$ is a spring constant, and $m_h$ is the mass of an atom. We assume pairwise interactions between S and R, and within R, so that $\Phi$ in \eqn{eqn:prob_2D} is given by
\begin{equation}\label{eqn:phi}
    \Phi(\bm{q}^{\rm S};T) =
   \int_{\Gamma_{\bm{q}^{\rm R}}}
\prod_{i}^{N_{\rm S}}
\prod_{j}^{N_{\rm R}}
\exp{\bigl[-\beta \phi(|\bm{q}_i^{\rm S}-\bm{q}_j^{\rm R}|)\bigr]}
    \prod_{k}^{N_{\rm R}}
    \prod_{l (l\neq k)}^{N_{\rm R}}
\exp{\bigl[-\beta \tilde\phi(|\bm{q}_k^{\rm R}-\bm{q}_l^{\rm R}|)/2\bigr]}
    d\bm{q}^{\rm R},
\end{equation}
where $\phi(r)$ is the interaction between an atom in R and an atom in S, and $\tilde\phi(r)$ is the interaction between two atoms in R. Note that $N_{\rm S}=2$ for the harmonic oscillator.
For the gas reservoir where gas--gas interactions ($\tilde\phi$) are negligible and all gas atoms are identical, \eqn{eqn:phi} takes the form,
\begin{align}\label{eqn:phi2}
    \Phi(\bm{q}^{\rm S};T) &=
    \int_{\Gamma_{\bm{q}^{\rm R}}}
\prod_{i}^{N_{\rm S}}
\prod_{j}^{N_{\rm R}}
\exp{\bigl[-\beta \phi(|\bm{q}_i^{\rm S}-\bm{q}_j^{\rm R}|)\bigr]}
    d\bm{q}^{\rm R}, \\ \nonumber
    &= \biggl[\int_{\Gamma_{\hat{\bm{q}}^{\rm R}}}
\prod_{i}^{N_{\rm S}}
\exp{\left[-\beta \phi(|\bm{q}_i^{\rm S}-\hat{\bm{q}}^{\rm R}|)\right]}
   d \hat{\bm{q}}^{\rm R}\biggr]^{N_{\rm R}} \\ \nonumber
    &= \biggl[\int_{\Gamma_{\hat{\bm{q}}^{\rm R}}}
\prod_{i}^{N_{\rm S}}
\exp{\left[-\beta \phi_i(\bm{q}_i^{\rm S})\right]}
   d \hat{\bm{q}}^{\rm R}\biggr]^{N_{\rm R}},
\end{align}
where $\phi_i\equiv\phi(|\bm{q}_i^{\rm S}-\bm{q}_j^{\rm R}|)$ is the interaction between atom $i$ in S and a generic gas atom in R located at position $\hat{\bm{q}}^{\rm R}$, where all $N_{\rm R}$ gas atoms are identical.
Using the ideal gas law ($N_{\rm R}=\beta p_{\rm R}V_{\rm R}$), \eqn{eqn:phi2} becomes
\begin{equation}\label{eqn:phi3}
     \Phi(\bm{q}^{\rm S};T,p_{\rm R}) = \biggl[\int_{\Gamma_{\hat{\bm{q}}^{\rm R}}}\prod_{i}^{N_{\rm S}}\exp{\bigl[-\beta \phi_{i}\bigr]}
    d\hat{\bm{q}}^{\rm R}\biggr]^{\beta p_{\rm R}V_{R}}.
\end{equation}
Substituting \eqns{eqn:H_har} and \eqnx{eqn:phi3} into
\eqn{eqn:prob_2D}, the PDF of the harmonic oscillator for $\bm{q}_1^{\rm S}$ and $\bm{q}_2^{\rm S}$ is
\begin{equation}\label{eqn:prob_harmonic}
    f_{\rm S}(\bm{q}_1^{\rm S},\bm{q}_2^{\rm S};T,p_{\rm R})
    = \frac{1}{\bar{Z}_{\rm S}} \exp{(-\beta \mathcal{V}_{\rm S})}
    \biggl[ \int_{\Gamma_{\hat{\bm{q}}^{\rm R}}} \exp{\bigl(-\beta (\phi_{1} +\phi_{2})\bigr)} d\hat{\bm{q}}^{\rm R}\biggr]^{\beta p_{\rm R} V_{\rm R}},
\end{equation}
where
\begin{equation}
    \bar{Z}_{\rm S} = \int_{\Gamma_{\bm{q}_1^{\rm S},\bm{q}_2^{\rm S}}}
    \exp{(-\beta \mathcal{V}_{\rm S})}
    \biggl[ \int_{\Gamma_{\hat{\bm{q}}^{\rm R}}} \exp{\bigl(-\beta (\phi_{1} +\phi_{2})\bigr)} d\hat{\bm{q}}^{\rm R}\biggr]^{\beta p_{\rm R} V_{\rm R}}
    d\bm{q}_1^{\rm S} d\bm{q}_2^{\rm S}.
\end{equation}
Here, we omit $\mathcal{K}_{\rm S}$ and $C(T)$ in \eqn{eqn:prob_harmonic} because they do not influence the form of the PDF in terms of the atom positions in S  ($\bm{q}_1^{\rm S}$ and $\bm{q}_2^{\rm S}$) (see the general form in \eqn{eqn:prob_2D}).
Since the distance between the two atoms in S ($r_{12}$) represents the degree of freedom of S, $f_{\rm S}\left(\bm{q}_1^{\rm S},\bm{q}_2^{\rm S};T_{\rm R},p_{\rm R}\right)$ in \eqn{eqn:prob_harmonic} is reformulated using spherical coordinate as follows:
\begin{equation}\label{eqn:prob_har}
    f_{\rm S}(r_{12};T_{\rm R},p_{\rm R}) = \frac{1}{\bar{Z}_{\rm S}}  r_{12}^2 \exp{(-\beta a r_{12}^2)}
    \Phi(r_{12};T_{\rm R},p_{\rm R})
    ,
\end{equation}
where
\begin{align}\label{eqn:prob_har_2}
 \Phi(r_{12};T_{\rm R},p_{\rm R}) &= \biggl[\int_{\Gamma_{\hat{\bm{q}}^{\rm R}}} \exp{\Bigl(-\beta (\phi_{1} +\phi_{2})\Bigr)} d\hat{\bm{q}}^{\rm R}\biggr]^{\beta p_{\rm R} V_{\rm R}}, \\
    \bar{Z}_{\rm S} &= \int_{\Gamma_{r_{12}}} r_{12}^2
    \exp{\bigl(-\beta a r_{12}^2\bigr)\Phi\bigl(r_{12};T_{\rm R},p_{\rm R}\bigr)} dr_{12}.
\end{align}
Note that we use $T_{\rm R}$ for $T$ because they are identical following the assumption in \sect{sec:stat_mec_2D}.

Given the harmonic oscillator spring constant $a$, reservoir properties $T_{\rm R}$ and $p_{\rm R}$, and a functional form for $\phi(r)$, $f_{\rm S}$ in \eqn{eqn:prob_har} can be numerically computed. We choose $a=0.01$~eV/\AA$^2$, $T_{\rm R}= 80$~K, and take $\phi(r)$ to be an LJ potential with parameters $\sigma = 5$~\AA\ and $\epsilon = 0.01$~eV with a cutoff radius of 12~\AA. The numerical integration in \eqn{eqn:prob_har} is carried out over a $\hat{\bm{q}}^{\rm R}$ domain spanning $\{(x,y,z)\in\mathbb{R}^3: -100 <x< 100\;\angstrom, -100 <y< 100\;\angstrom,\; \text{and} -100 <z< 100\;\angstrom\}$. This integration is computed using the \texttt{integral3()} method in MATLAB. For tested $r_{12}$ values, we verify that if the domain of $\hat{\bm{q}}^{\rm R}$ is large enough to encompass two spheres located at $\bm{q}_1^{\rm S}$ and $\bm{q}_2^{\rm S}$ with radii equal to the cutoff distance,  the domain size of $\hat{\bm{q}}^{\rm R}$ does not affect the PDF $f_{\rm S}$ in \eqn{eqn:prob_har}.

\begin{figure}\centering
    \includegraphics[width=0.7\textwidth]{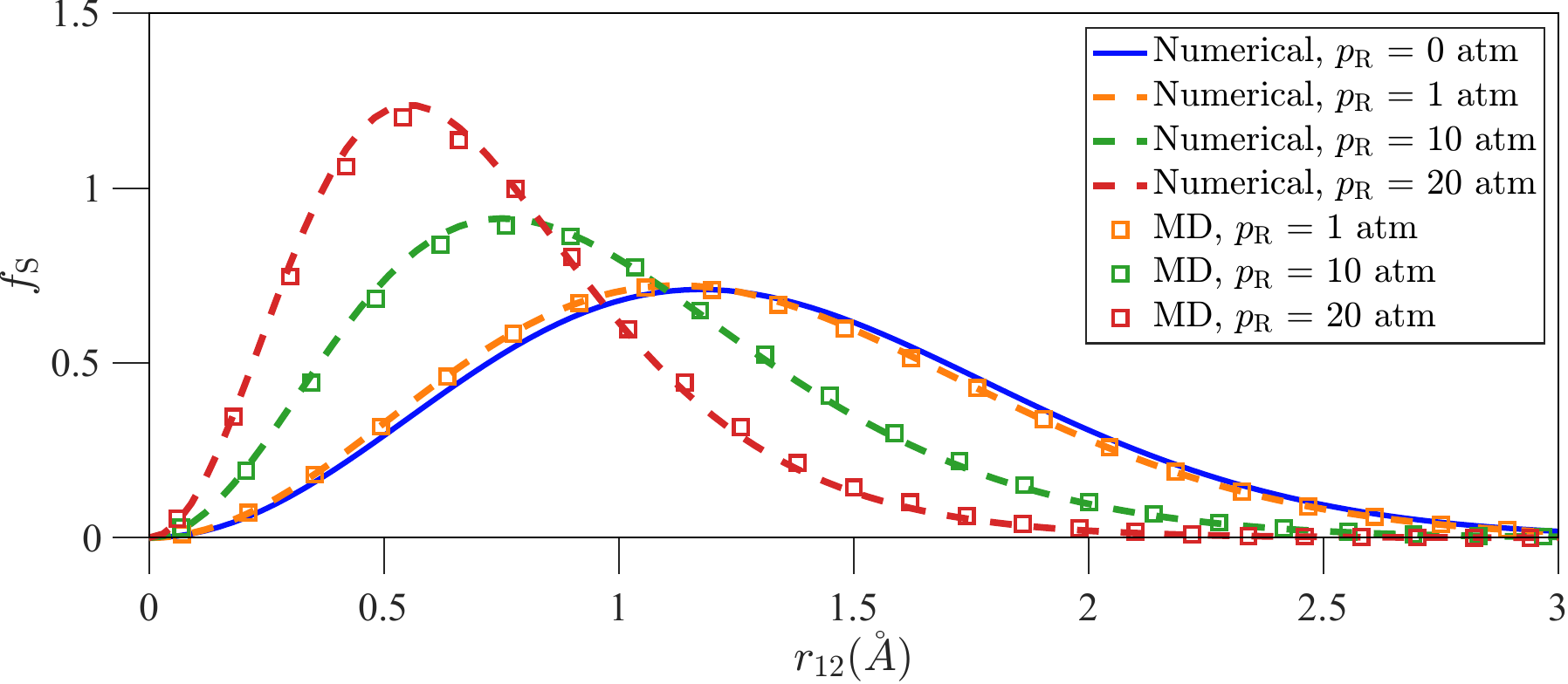}
    \caption{Computed $f_{\rm S}(r_{12})$ (\eqn{eqn:prob_har}) for different pressures $p_{\rm R}$.
    The solid blue line represents $f_{\rm S}(r_{12})$ following \eqn{eqn:prob} with $p_{\rm R}=0$ atm (i.e., $\Phi=1$), dashed lines represent the numerically computed $f_{\rm S}(r_{12})$ PDFs, and the squares represent the measured $f_{\rm S}(r_{12})$ from the MD simulation for different pressures $p_{\rm R}$.}
    \label{fig:f_r12}
\end{figure}

For comparison, we perform MD simulations for the same conditions using the gas reservoir thermostat of \sect{sec:gas_res_method}. In simulations, the harmonic oscillator is positioned at the center of the simulation box. The mass of all atoms is taken to be 12.01~u.\footnote{Since the mass of the harmonic oscillator and gas atoms does not influence $f_{\rm S}$, we set this value arbitrarily.} The simulation box size is $200\times200\times300\;\angstrom^3$ with two gas generation surfaces positioned at $z=$ 50 and 250~\AA.
The LJ parameters for the interaction between a gas atom and a harmonic oscillator atom,  and the spring constant, are the same as those used for the numerical integration.
Note that the interaction between gas atoms is ignored so as to obey the assumptions used in the derivation of \eqn{eqn:prob_har}.
One of the atoms in the harmonic oscillator is fixed at the center of the gas reservoir to prevent rigid-body translation. $T_{\rm R}$ is set to 80~K for all pressure cases. The simulation is run for 200~ns to reach equilibrium for several different pressures ($p_{\rm R}=1$, 10, and 20~atm), followed by an additional 100~ns to record $r_{12}$ with a sampling interval of 1~ps. The sampled $r_{12}$ values are used to compute the $f_{\rm S}(r_{12})$ distribution.

\begin{figure}\centering
\includegraphics[width=0.5\textwidth]{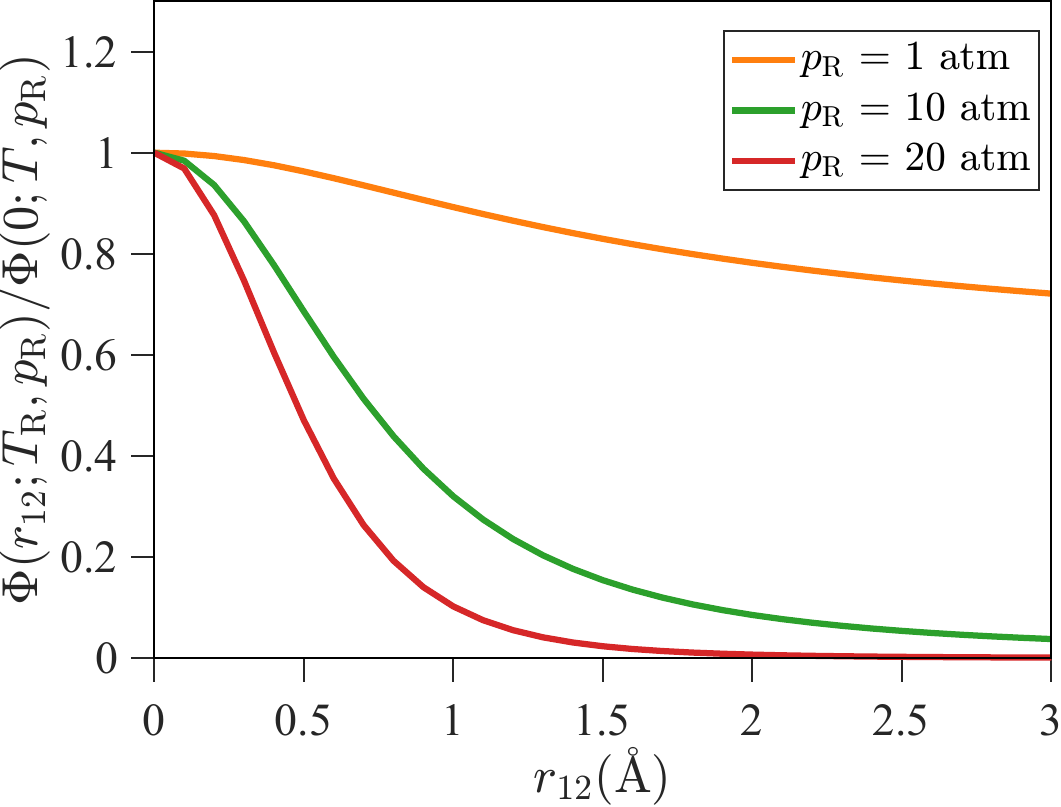}
    \caption{Numerically computed $\Phi(r_{12};T_{\rm R},p_{\rm R})$ for different pressures $p_{\rm R}$ (1, 10, 20 atm) at $T_{\rm R}$ = 80 K. $\Phi(r_{12};T_{\rm R},p_{\rm R})$ is normalized by $\Phi(0;T_{\rm R},p_{\rm R})$.}
    \label{fig:phi_P}
\end{figure}

The results from the numerical integration and MD simulations are compared in \fig{fig:f_r12}.  Distributions of $r_{12}$ from both methods are in excellent agreement across all pressures. For $p_{\rm R}=0$~atm, \eqn{eqn:prob_har} becomes the PDF of the harmonic oscillator in the conventional canonical ensemble (\eqn{eqn:prob}) since $\Phi=1$. As $p_{\rm R}$ increases, $f_{\rm s}$ becomes more narrow with the peak shifted left, indicating a higher probability at lower $r_{12}$ values.
This shift in distribution with increasing pressures occurs because $\Phi$ reaches its maximum value when the two atoms are at the same position (i.e., $r_{12}$=0) as shown in \fig{fig:phi_P}. In \eqn{eqn:prob_har_2}, the integration in $\Phi$ is raised to the power of $N_{\rm R}=p_{\rm R}V_{\rm R}/(k_{\rm B} T_{\rm R})$. As $p_{\rm R}$ increases at constant $T_{\rm R}$, $N_{\rm R}$ increases proportionally, raising the same integration in $\Phi$ to the power of the increased $N_{\rm R
}$. This exponentiation leads to the observed narrowing of the $f_{\rm S}$ distribution in \fig{fig:f_r12}.

\section{Application to a Au nanoparticle}\label{sec:np_eq}
\begin{figure}[t]\centering
    \includegraphics[width=1.0\textwidth]{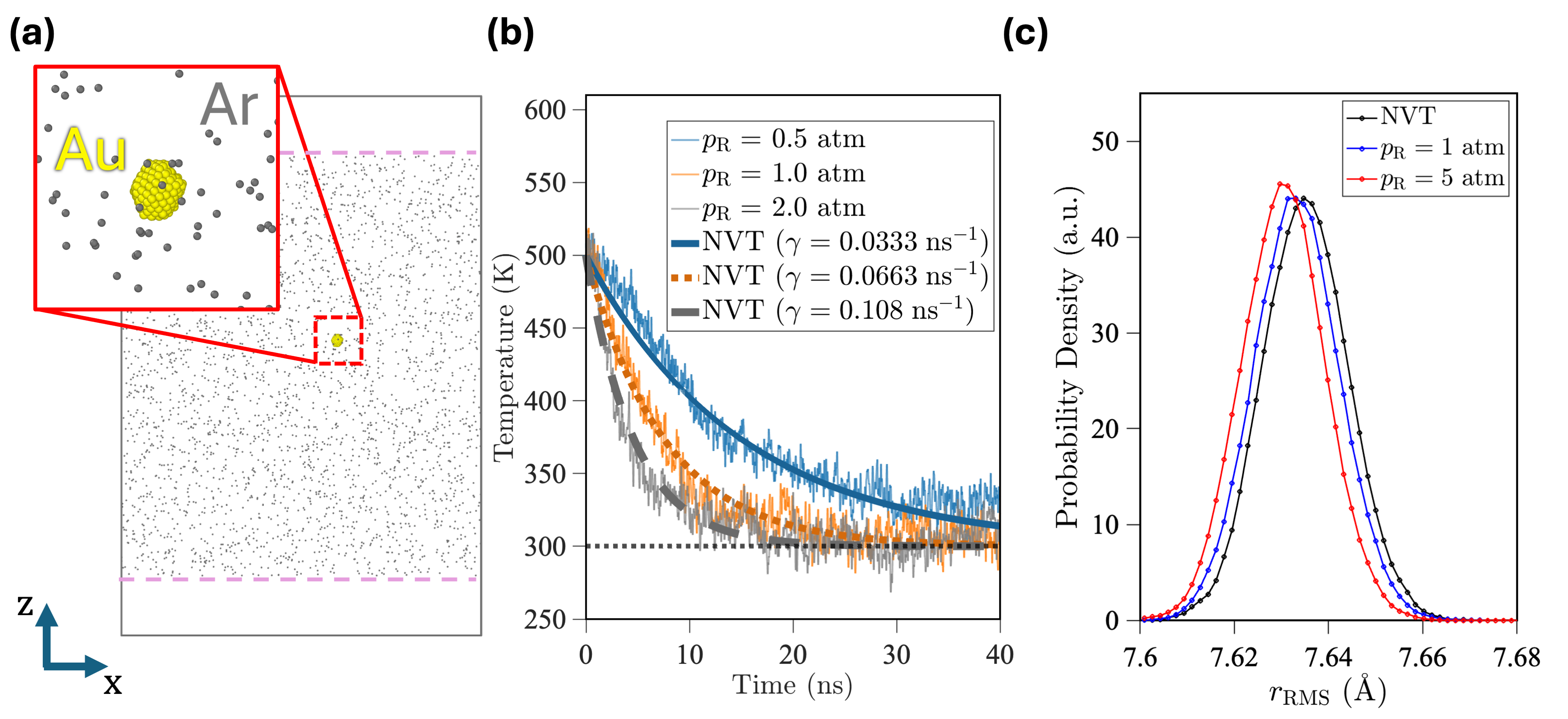}
\caption{Equilibration of a Au nanoparticle in an Ar gas reservoir.
(a) Atomistic configuration of a Au nanoparticle embedded in an Ar gas reservoir; purple dashed lines indicate the reservoir boundaries.
(b) Time evolution of the nanoparticle temperature for reservoir pressures $p_{\rm R}=1$, 2, and 3~atm. The nanoparticle cools from the initial temperature $T_{\rm init}=500$~K toward the reservoir temperature $T_{\rm R}=300$~K. Thick curves show the analytical temperature relaxation predicted by Langevin dynamics, with damping coefficients fitted separately for each $p_{\rm R}$.
(c) Probability density distributions of the RMS radius ($r_{\rm RMS}$)
of a Au nanoparticle for the NVT thermostat and the gas-reservoir at $p_{\rm R}=1$~atm and $p_{\rm R}=5$~atm.}
\label{fig:np_eq}
\end{figure}

To go beyond an idealized 1D harmonic oscillator, we consider a Au nanoparticle within an Ar gas reservoir, and compare it with thermostat-based equilibration.
\fig{fig:np_eq}a shows the atomistic simulation setup. The  nanoparticle contains 242 atoms and is modeled using an embedded-atom method (EAM) potential \cite{aueam}, while Ar--Au and Ar--Ar interactions are modeled via LJ potentials (\eqn{eqn:LJ}) with parameters: $\epsilon_{\rm ArAu}=0.0395$~eV, $\sigma_{\rm ArAu}=3.17$~\AA\ \cite{heinz2013}, and $\epsilon_{\rm ArAr}= 0.0103~$eV, $\sigma_{\rm ArAr}=3.4$~\AA\ \cite{heinz2013}. The simulation box size is $600 \times 300 \times 900~\text{\AA}^3$ with the gas surfaces located at $z=100$~\AA~and $z=800$~\AA.

Initially, the Au nanoparticle at the center of the simulation box is thermalized to $T_{\rm init}=500$~K by applying a Langevin thermostat solely to it for 100 ps.
Next, the nanoparticle is frozen while an Ar gas reservoir at $T_{\rm R}=300$~K is initialized; during this stage, Ar--Au interactions are disabled. After 1~ns, once the gas atoms have filled the reservoir, any Ar atom within 5~\AA\ of a nanoparticle atom is removed to prevent close collisions, and the nanoparticle is released with atom velocities set equal to their values at the end of the initial thermalization stage. Next, Ar--Au interactions are turned on, allowing thermalization to proceed via gas--nanoparticle collisions. The vibrational temperature of the nanoparticle is computed every 10~ps after removing translational and rotational kinetic energy contributions to the temperature.

\fig{fig:np_eq}b shows the time evolution of the nanoparticle temperature after the Ar–Au interactions are activated for three representative reservoir pressures, $p_{\rm R}=0.5$, 1, and 2 atm. In all cases, the nanoparticle temperature equilibrates toward the reservoir temperature of 300~K. Increasing $p_{\rm R}$ leads to faster equilibration, as the higher gas density results in more frequent gas–nanoparticle collisions and enhanced energy exchange.
For comparison, the thick lines in \fig{fig:np_eq}b show the analytical temperature relaxation expected from Langevin dynamics (see Section S.2 in the SI for the derivation),
\begin{equation}\label{eqn:damping_eq}
T(t)=T_{\rm target}+\bigl(T_{\rm init}-T_{\rm target}\bigr)\exp(-2\gamma t),
\end{equation}
where $T_{\rm target}=300$~K is a target temperature, and $\gamma$ is the damping coefficient fitted separately for each reservoir pressure $p_{\rm R}$.
The results indicate that the reservoir pressure $p_{\rm R}$ plays a role analogous to the damping coefficient in a Langevin thermostat, i.e., it controls how rapidly the nanoparticle temperature relaxes toward equilibrium. Higher pressures correspond to stronger effective damping, leading to faster thermal equilibration through more frequent gas–nanoparticle collisions.

The results above show that the presence of the gas affects the transient response of the nanoparticle prior to equilibration. It is also of interest to study the effect on equilibrium properties. As such we analyze the probability distribution of the nanoparticle effective size defined by its root-mean-square radius ($r_{\rm RMS}$). At a given time $t$, $r_{\rm RMS}$ is defined as
\begin{equation}
r_{\rm RMS}(t) = \sqrt{\frac{1}{N}\sum_{i=1}^{N} \left| \mathbf{r}_i(t) - \mathbf{r}_{\rm COM}(t) \right|^2 },
\end{equation}
where $\mathbf{r}_i$ is the position of atom $i$, $\mathbf{r}_{\rm COM}$ is the center-of-mass position of the nanoparticle, and $N$ is the total number of Au atoms. Figure~\ref{fig:np_eq}c compares the probability density distributions of $r_{\rm RMS}$ obtained from an NVT simulation employing a Langevin thermostat ($\gamma=0.1~\rm{ns}^{-1}$)\footnote{The distribution of $r_{\rm RMS}$ is an equilibrium property and is therefore independent of the Langevin thermostat damping coefficient.} and from gas-reservoir simulations at $p_{\rm R}=1$~atm and $5$~atm. We observe systematic shifts in the distributions depending on the equilibration method and reservoir pressure. The mean $r_{\rm RMS}$ values are $7.635$~\AA\ for the NVT case, $7.633$~\AA\ for $p_{\rm R}=1$~atm, and $7.630$~\AA\ for $p_{\rm R}=5$~atm. The gradual decrease in the mean $r_{\rm RMS}$ with increasing pressure indicates a compression of the nanoparticle induced by surrounding gas atoms, deviating from the structural statistics expected in an NVT ensemble. The effect is small for the studied pressures, but becomes more pronounced as $p_{\rm R}$ increases reflecting the enhanced frequency and momentum transfer of gas–nanoparticle collisions.\footnote{At higher pressures, the Ar atoms layer onto the nanoparticle, leading to a qualitatively different behavior not included here.} In this regime, the $\Phi$ term in \eqn{eqn:prob_2D}, which accounts for the direct coupling between the nanoparticle and the surrounding gas, can no longer be neglected.
In contrast to conventional thermostat-based approaches, which implicitly assume negligible system-bath coupling forces, the gas-reservoir method explicitly incorporates both energy and momentum exchange between the system and heat bath. As a result, it provides a more physically faithful representation of thermal and structural equilibrium.

\section{Application to a 2D Graphene monolayer}\label{sec:gra_prop}
The interaction between a 2D material and gas atoms is inevitable in an experimental setup unless the 2D material is placed in a complete vacuum. As discussed in \sect{sec:stat_mec_2D}, this interaction influences the PDF of the 2D material, potentially altering its properties. To investigate how the properties of a 2D material are affected by a gaseous environment, we consider  a graphene monolayer in a gas reservoir. Specifically, we examine the out-of-plane fluctuation of the graphene, which is governed by the PDF of the instantaneous states of the system within the ensemble. The computed fluctuations are then compared with those obtained from the conventional canonical (NVT) ensemble, where the PDF is defined by \eqn{eqn:prob}.

\subsection{Method}
\label{sec:gra_prop:method}
The standard deviation of the out-of-plane fluctuation of a graphene monolayer is given by
\begin{equation}\label{eqn:h_std}
    h_{\rm std} = \sqrt{\frac{1}{N}\sum_{i=1}^{N}\left(z_i-\bar{z}\right)^2},
\end{equation}
where $\bar{z}$ is given by
\begin{equation}
    \bar{z} = \frac{1}{N}\sum_{i=1}^{N}z_i.
\end{equation}
Here, $N$ is the number of carbon atoms in the graphene monolayer, and $z_i$ is the $z$ position (in the out-of-plane direction) of atom $i$.
Using the PDF in \eqn{eqn:prob_2D}, the average value of $h_{\rm std}$ for the graphene monolayer in the gas reservoir is
\begin{equation}\label{eqn:gas:h_std_avg}
    \langle h_{\rm std}\rangle = \frac{1}{Z_{\rm S}}\int_{\Gamma_{\bm{q}^{\rm S}}}  h_{\rm std} \exp{(-\beta \mathcal{V}_{\rm S})}\Phi(\bm{q}^{\rm S};T_{\rm R},p_{\rm R})d\bm{q}^{\rm S},
\end{equation}
where
\begin{equation}
    Z_{\rm S} = \int_{\Gamma_{\bm{q}^{\rm S}}}\exp{(-\beta \mathcal{V}_{\rm S})}\Phi(\bm{q}^{\rm S};T_{\rm R},p_{\rm R}) d\bm{q}^{\rm S}.
\end{equation}
In contrast to the harmonic oscillator in \sect{sec:num_harmonic}, evaluating \eqn{eqn:gas:h_std_avg} using numerical integration is challenging due to the large number of degrees of freedom in the graphene system. Instead, $\langle h_{\rm std}\rangle$ in \eqn{eqn:gas:h_std_avg} is computed as a time average over the trajectory in the MD gas reservoir simulation,
\begin{equation}\label{eqn:gas:h_std_MD}
    \bar{h}_{\rm std} = \frac{1}{N_{\rm sampling}}\sum_{j=1}^{N_{\rm sampling}} h_{\rm std}(t_j),
\end{equation}
where $N_{\rm sampling}$ is the total number of simulation time steps for sampling, and $h_{\rm std}(t_j)$ is $h_{\rm std}$ at time $t_j$.

\begin{figure}[t]\centering
\includegraphics[width=0.35\textwidth]{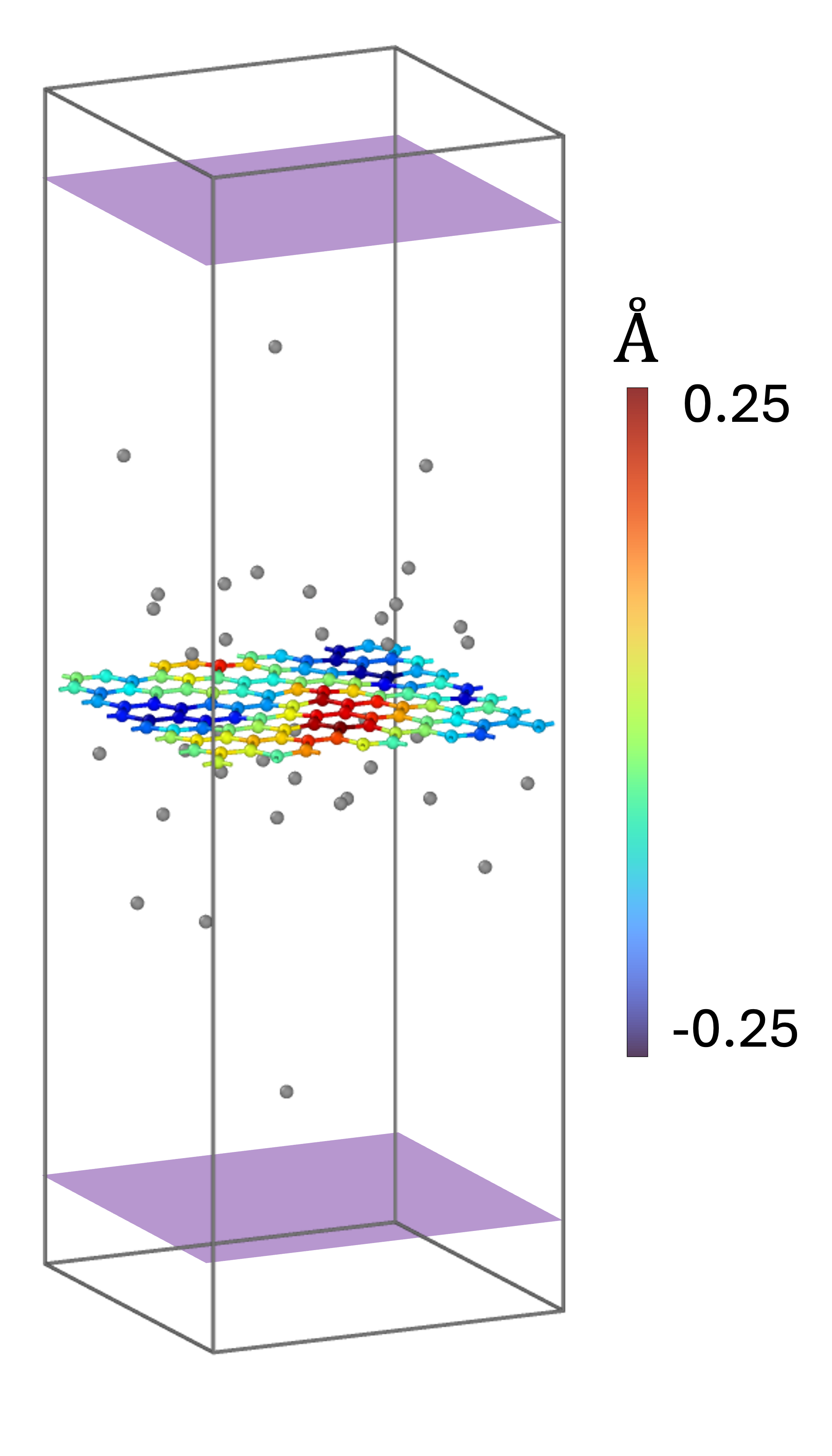}
    \caption{Snapshot of a graphene monolayer in the gas reservoir MD simulation. The colormap represents the out-of-plane displacement relative to the center of mass ($z_i-\bar{z}_i$ in \eqn{eqn:h_std}) of the graphene atoms. Gas atoms in the reservoir are shown in gray.}
    \label{fig:graphene_gas}
\end{figure}

The rectangular graphene monolayer is created using the Nanomaterial Modeler \cite{choi2021} in CHARMM-GUI \cite{jo2008}, and interactions in the graphene monolayer are modeled using the interface force field (IFF) \cite{pramanik2017}.

\fig{fig:graphene_gas} illustrates the MD simulation setup for the graphene monolayer in the gas reservoir. The graphene monolayer has dimensions of $19.148\times16.58\; \angstrom^2$ (128 carbon atoms).  The size of the simulation box for the gas reservoir is $19.148\times16.58\times80$ \AA$^3$, with two gas generation surfaces located at $z=20$ and $60\;\angstrom$. The graphene monolayer is positioned at $z = 40\;\angstrom$. PBCs are applied in the in-plane directions to remove edge effects and prevent rotation of the graphene within the simulation box.
Argon atoms are selected for the gas reservoir simulation, utilizing the same LJ parameters and mass used in \sect{sec:gas_res_num}. Note that gas--gas interactions are considered in this simulation in contrast the simulations in \sect{sec:num_harmonic}. The LJ parameters for the cross-species (C--Ar) interaction are $\epsilon=0.0081$~eV and $\sigma=3.51~\angstrom$ \cite{heinz2013}.
To prevent vertical translation of the graphene monolayer, a spring is connected to its center of mass along the z direction using the \texttt{fix spring} method in LAMMPS. The spring constant is 10.0 eV/\AA$^2$. We confirm that the spring constant value does not affect the out-of-plane fluctuation results in the simulations.

\subsection{MD simulation results}
MD simulations of the graphene monoloayer in the gas reservoir are performed at various conditions, $T_{\rm R}=250$, 300, 350, 400, 450 and 500~K, and $p_{\rm R}=5$, 10, and 20~atm. $\bar{h}_{\rm std}$ is measured for each of these cases according to \eqn{eqn:gas:h_std_MD}. In all cases, $N_{\rm sampling} = 2.0 \times 10^{5}$, corresponding to sampling every 0.5~ps over a total sampling duration of 100~ns after equilibration. For comparison, MD simulations are also performed for a graphene monolayer without a surrounding gas using a Langevin thermostat to maintain temperature, thereby following the conventional canonical (NVT) ensemble in \eqn{eqn:prob}. The standard deviation obtained in these simulations is denoted $\bar{h}_{\rm std}^{\rm NVT}$. The damping coefficient for the Langevin thermostat is set to 1 ps.
Simulations are conducted for 100 ns to equilibrate the monolayer graphene followed by an additional 100 ns of simulation, while the positions of carbon atoms are sampled every 0.1 ps.

\begin{figure}[t]\centering
\includegraphics[width=0.7\textwidth]{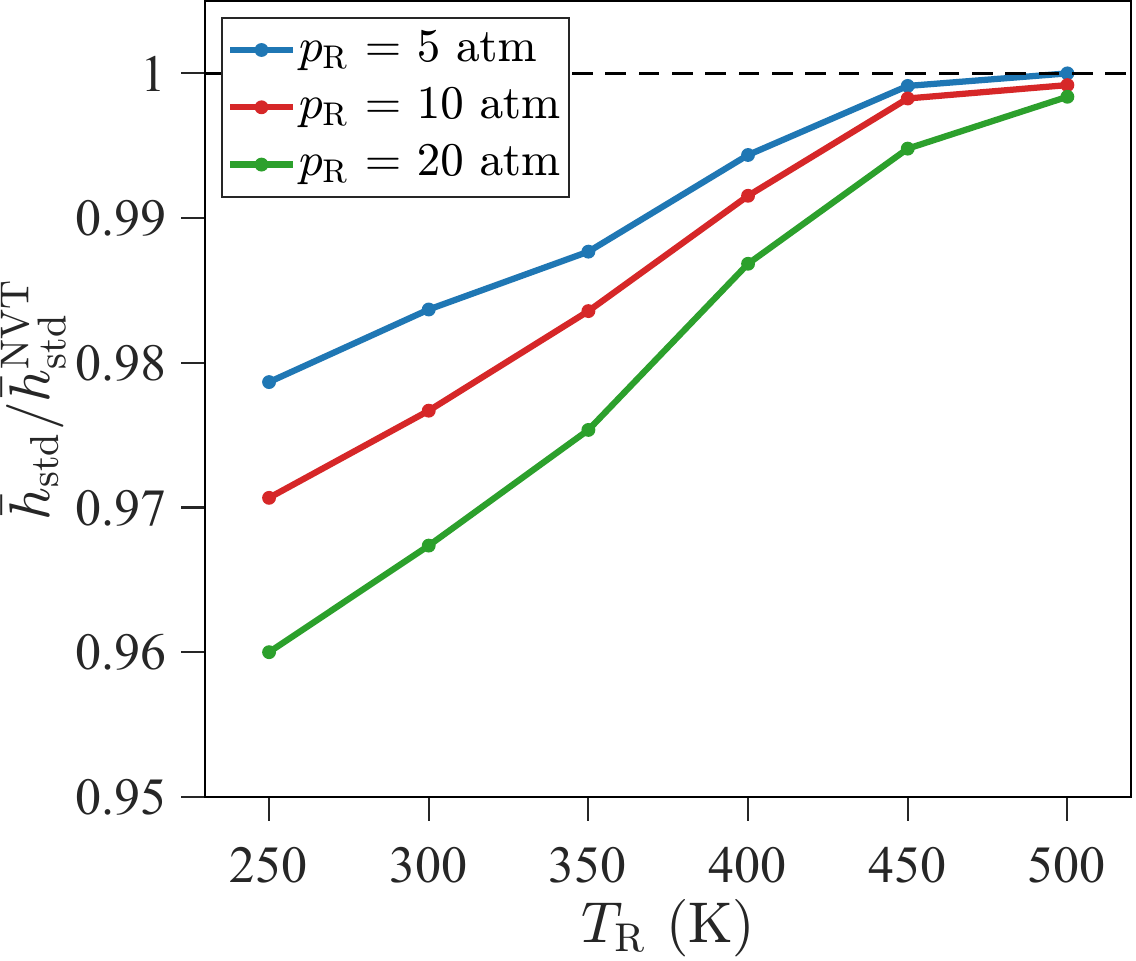}
    \caption{Standard deviation in out-of-plane fluctuations in monolayer graphene in a gas reservoir as a function of the reservoir temperature for different pressures, normalized by the canonical ensemble results obtained using a Langevin thermostat.}
    \label{fig:h_std}
\end{figure}

\fig{fig:h_std} shows the measured $\bar{h}_{\rm std}$ from the gas reservoir simulations as a function of $T_{\rm R}$ for different values of $p_{\rm R}$, normalized by $\bar{h}_{\rm std}^{\rm NVT}$ to highlight how the introduction of a gas reservoir causes deviations from the NVT ensemble reference value. The deviations arise due to the non-negligible interaction between the gas and graphene, as well as between the gas atoms themselves. As the reservoir temperature increases at a fixed pressure, the number density of gas atoms decreases, leading to fewer gas–graphene collisions. This reduction in collision frequency results in a weak interaction with the graphene monolayer. As a consequence, the graphene dynamics follow the canonical ensemble statistics, and $\bar{h}_{\rm std}/\bar{h}_{\rm std}^{\rm NVT}$ converges to unity.
At the other extreme we see that $\bar{h}_{\rm std}$/$\bar{h}_{\rm std}^{\rm NVT}$ decreases with decreasing temperature.\footnote{
When the temperature falls below $T_{\rm R}\approx250$~K, Ar atoms are deposited onto the graphene monolayer due to reduced thermal motion and strong gas--graphene interactions. This continued adsorption leads to the growth of dense Ar layers on the graphene surface, which rapidly fills the simulation box.}  This effect intensifies with increasing $p_{\rm R}$ because the number of gas atoms in the gas reservoir---and consequently the gas--graphene interactions---increases.

To assess the reliability of our results, we compute the 95\% confidence interval using $1.98/h^*_{\rm std} \sqrt{N_{\rm sample}}$ where  $h_{\rm std}^*$ represents the standard deviation of sampled $h_{\rm std}$ during  simulations. The resulting confidence interval is less than 0.1\% for both $\bar{h}_{\rm std}$ and $\bar{h}_{\rm std}^{\rm NVT}$,
indicating that the reported deviations are statistically significant and not due to sampling noise.

As discussed in \sect{sec:gra_prop:method}, performing the numerical integration to obtain the PDF of the graphene within the gas reservoir is challenging due to the number of degrees of freedom involved. However, the harmonic oscillator results presented in \sect{sec:num_harmonic} provide useful physical insight: interactions with the surrounding gas bias the PDF toward configurations with reduced internal fluctuations, as evidenced by the increased probability of smaller $r_{12}$ in \fig{fig:f_r12}. By analogy, gas--graphene interactions suppress out-of-plane fluctuation of the graphene monolayer because such configurations increase its instantaneous surface area and thereby enhance interactions with the surrounding gas. Consequently, configurations with smaller out-of-plane fluctuations and reduced effective surface area are statistically favored. As the number of gas atoms in the reservoir increases, this constraint becomes stronger, leading to further suppression of out-of-plane motion and resulting in $\bar{h}_{\rm std} / \bar{h}_{\rm std}^{\rm NVT} < 1$.

\section{Conclusion}
\label{sec:conclusion}
A PDF is derived for a system in the canonical ensemble that explicitly accounts for the interactions with a heat bath (typically neglected under the assumptions of ``weak interactions''). To validate this PDF, an MD algorithm is developed for simulating a system in contact with an infinite heat bath by introducing an intermediary gas reservoir in which the gas atoms are modeled explicitly. Evaluation of the pressure, temperature, and atom velocities of the gas reservoir confirm that the simulation reproduces equilibrium statistical mechanics of an ideal gas.

The PDF is computed and validated using MD simulations for representative nanoscale systems where the weak interaction assumption is suspect: a 1D harmonic oscillator, a Au nanoparticle, and a 2D graphene monolayer. In all cases, the MD results are consistent with the predictions of the theory. For the 1D harmonic oscillator, an explicit expression is obtained for the PDF that is numerically integrated and found to be in excellent agreement with MD simulations over a range of gas reservoir pressures. The results show a strong dependence of the oscillator's probable length on reservoir pressure—an effect that is missing in the conventional canonical ensemble. For the Au nanoparticle, we show that the thermal relaxation dynamics depend on the reservoir pressure. After equilibration, we find that its effective radius exhibits deviations from canonical predictions as the pressure increases. Similarly, for the 2D graphene monolayer, results indicate deviations in out-of-plane fluctuations compared to those predicted by a conventional canonical ensemble simulation of the monolayer coupled to a Langevin thermostat.

The derived PDF reveals the influence of environmental factors on the properties of low-dimensional materials, suggesting several promising directions for future research:
\begin{itemize}
\item Substrate effect: Since 2D materials typically require a supporting substrate, the influence of the substrate on the PDF of the 2D material is substantial, exceeding that of a gas reservoir due to the higher number density of the substrate. It is of interest to study the effect of different substrates on 2D material properties through the lens of equilibrium statistical mechanics including weak interactions as presented here.
\item Liquid environment: Our findings show that the effect of the gas reservoir on 2D materials increases with pressure. This suggests extending this study to a liquid environment where the higher density of atoms compared to a gas could result in stronger interactions with the 2D material.
\end{itemize}

Beyond the theory, the gas reservoir algorithm developed here provides an efficient and physically grounded approach for performing MD simulations of a system in contact with a heat bath. It is well known that conventional thermostatting techniques, such as Langevin and Nos\'e--Hoover methods, can significantly distort the intrinsic dynamics of the system \cite{basconi2013}. Such issues are particularly pronounced in low-dimensional and nanoscale systems, where the choice of thermostatting strategy has been shown to strongly influence transport properties \cite{guvensoy2024}. In contrast, the ``gas reservoir thermostat'' introduced in this work yields the correct equilibrium behavior through explicit, physically meaningful interactions with an environment, while simultaneously accounting for pressure effects of the surrounding gaseous medium. Therefore, this approach offers a promising alternative for a wide range of low-dimensional materials and surface chemistry problems, where conventional thermostat-based simulations may lead to artifacts in both dynamics and structural properties.

\section*{Acknowledgments}
The authors thank Prof.\ Tom Schwartzentruber and Dr.\ Erik Torres (University of Minnesota) for helpful discussions on the gas reservoir simulation algorithm, and Prof.\ Chris Hogan, Aayushi Bohrey, and Georgia Robles (University of Minnesota) for their assistance with the Au nanoparticle simulations.

\printbibliography

\end{document}


\maketitle
\renewcommand{\thesection}{S.\arabic{section}}
\renewcommand{\thesubsection}{S.\arabic{section}.\arabic{subsection}}
\renewcommand{\thesubsubsection}{S.\arabic{section}.\arabic{subsection}.\arabic{subsubsection}}
\setcounter{section}{0}
\renewcommand{\theequation}{S\arabic{equation}}

\section{Equipartition and virial theorems in the presence of interactions with a heat bath}\label{sec:equip}
Following the approach in \cite{tadmor2011}, we derive the equipartition and virial theorems for the PDF in \eqn{eqn:prob_2D} in the main text, which accounts for the weak interactions with a heat bath that are neglected in the conventional canonical ensemble. The ensemble average of $x_i(\partial\mathcal{H}_{\rm S}/\partial x_j)$, where $\bm{x}$ is either the set of position $\bm{q}$ or the set of momenta $\bm{p}$, is given by
\begin{equation}\label{eqn:equi_general}
\langle x_i \frac{\partial\mathcal{H}_{\rm S}}{\partial x_j}  \rangle = \frac{C(T)}{Z_{\rm SR}}\int_{\Gamma_{\bm{q}^{\rm S},\bm{p}^{\rm S}}} x_i \frac{\partial \mathcal{H}_{\rm S}}{\partial x_j} \exp(-\beta \mathcal{H}_S)\Phi d\bm{q}^{\rm S}d\bm{p}^{\rm S},
\end{equation}
where we omit dependencies of $\mathcal{H}$ and $\Phi$ for brevity. Using the identity,
\begin{equation}
\frac{\partial}{\partial x_j}\Bigl(x_i \exp(-\beta\mathcal{H}_{\rm S})\Phi\Bigr)=
\delta_{ij}\exp(-\beta\mathcal{H}_S)\Phi-\frac{1}{k_BT}x_i\frac{\partial \mathcal{H}_{\rm S}}{\partial x_j}\exp(-\beta\mathcal{H}_{\rm S})\Phi+x_i\exp(-\beta\mathcal{H}_{\rm S} )\frac{\partial\Phi}{\partial x_j},
\end{equation}
\eqn{eqn:equi_general} is rewritten as
\begin{equation}\label{eqn:equi_avg}
\langle x_i \frac{\partial\mathcal{H_{\rm S}}}{\partial x_j}  \rangle =   k_BT\bigl[\delta_{ij} + \frac{C(T)}{Z_{\rm SR}}(-I_1 + I_2)\bigr],
\end{equation}
where
\begin{align}
I_1 &=  \int_{\Gamma_{\bm{q}^{\rm S},\bm{p}^{\rm S}}} \frac{\partial}{\partial x_j}\bigl(x_i\exp(-\beta\mathcal{H}_{\rm S})\Phi\bigr) d\bm{q}^{\rm S}d\bm{p}^{\rm S}, \\
I_2 &=  \int_{\Gamma_{\bm{q}^{\rm S},\bm{p}^{\rm S}}} x_i\exp(-\beta \mathcal{H}_{\rm S})\frac{\partial \Phi}{\partial x_j} d\bm{q}^{\rm S}d\bm{p}^{\rm S}.
\end{align}
With $\bm{x}=\bm{p}$, $I_1$ is expressed as
\begin{align}
I_1 &= \int_{\Gamma_{\bm{q}^{\rm S}}}\exp(-\beta\mathcal{V}_{\rm S})\Phi
d\bm{q}^{\rm S}
\int_{\Gamma_{\bm{p}^{\rm S}}}\frac{\partial}{\partial p_j}\bigl(p_i\exp(-\beta\mathcal{K}_{\rm S})\bigr)d\bm{p}^{\rm S} \\ \nonumber
&= \int_{\Gamma_{\bm{q}^{\rm S}}}\exp(-\beta\mathcal{V}_S)\Phi
d\bm{q}^{\rm S}\int\limits_{-\infty}^{\infty}\cdots\int\limits_{-\infty}^{\infty}\Biggl[\int\limits_{-\infty}^{\infty}\frac{\partial}{\partial p_j}\bigl(p_i \exp(-\beta\mathcal{K}_{\rm S})\bigr)\Biggr]\ldots dp^{\rm S}_{j-i}dp^{\rm S}_{j+i}\ldots \\ \nonumber
&= \int_{\Gamma_{\bm{q}^{\rm S}}}\exp(-\beta\mathcal{V}_S)\Phi
d\bm{q}^{\rm S}\int\limits_{-\infty}^{\infty}\cdots\int\limits_{-\infty}^{\infty}\biggl[p_i \exp(-\beta\mathcal{K}_{\rm S})\biggr]\Biggl|_{p_j=-\infty}^{\infty}\ldots dp^{\rm S}_{j-i}dp^{\rm S}_{j+i}\ldots.
\end{align}
Here, $[\cdot]|$ is zero, leading to $I_1=0$. We refer Section 7.4.4 in \cite{tadmor2011} for details. $I_2$ is also zero because $\partial\Phi/\partial p_j$ in $I_2$ is zero.
Therefore, \eqn{eqn:equi_avg} simplifies to the equipartition theorem,
\begin{equation}\label{eqn:equi_moment}
\langle p_i \frac{\partial\mathcal{H}_{\rm S}}{\partial p_j}  \rangle  = \delta_{ij}k_B T.
\end{equation}
\Eqn{eqn:equi_moment} is identical to that of the conventional canonical ensemble \cite{tadmor2011}.

With $\bm{x}=\bm{q}$, $I_1$ becomes
\begin{equation}
I_1 = \int_{\Gamma_{\bm{q}^{\rm S}}}\frac{\partial}{\partial q_j}\bigl(q_i\exp(-\beta\mathcal{V}_S)\Phi\bigr)
d\bm{q}^{\rm S}
\int_{\Gamma_{\bm{p}^{\rm S}}}\exp(-\beta\mathcal{K}_S)d\bm{p}^{\rm S}.
\end{equation}
The first integration in $I_1$ is zero, following the approach explained in Section 7.4.4 of \cite{tadmor2011}. Therefore, we have $I_1=0$. $I_2$ is given by
\begin{equation}
I_2 = \int_{\Gamma_{\bm{q}^{\rm S}}} q_i\exp(-\beta \mathcal{V}_{\rm S})\frac{\partial \Phi}{\partial q_j} d\bm{q}^{\rm S}
\int_{\Gamma_{\bm{p}^{\rm S}}}\exp(-\beta\mathcal{K}_S)d\bm{p}^{\rm S}.
\end{equation}
In contrast to $I_1=0$, $I_2$ is non zero. Consequently the virial theorem is
\begin{equation}\label{eqn:equi_position}
\langle q_i \frac{\partial\mathcal{H_{\rm S}}}{\partial q_j}  \rangle  = k_BT \Biggl(\delta_{ij}+  \frac{\int_{\Gamma_{\bm{q}^{\rm S}}} q_i\exp(-\beta \mathcal{V}_{\rm S})\frac{\partial \Phi}{\partial q_j} d\bm{q}^{\rm S}}{\int_{\Gamma_{\bm{q}^{\rm S},\bm{q}^{\rm R}}}\exp\bigl(-\beta(\mathcal{V}_{\rm S}+\mathcal{V}_{\rm SR}+\mathcal{V}_{\rm R})\bigr)d\bm{q}^{\rm S}d\bm{q}^{\rm R}}\Biggr).
\end{equation}
This differs from the conventional canonical ensemble that has the same right-hand side as the equipartition theorem in \eqn{eqn:equi_moment}. As a result, $\langle q_i( \partial\mathcal{H}_{\rm S}/\partial q_j) \rangle$ can be a non-zero when $i\neq j$, and even for $i=j$, it is influenced by the complex interactions between the atoms in the system S and the surrounding gas reservoir R through the additional term in \eqn{eqn:equi_position}.

As an example, we evaluate the equipartition and virial theorems for the 1D harmonic oscillator discussed in \sect{sec:num_harmonic}, using the
interatomic distance $r_{12}$ as the sole configurational degree of freedom.
The 1D harmonic oscillator's hamiltonian is
\begin{equation}\label{eqn:H_har_2}
    \mathcal{H}_{\rm S} = \mathcal{V}_{\rm S} + \mathcal{K}_{\rm S} =a(r_{12})^2 + \sum_{i=1}^2\frac{1}{2}m_h\vert\bm{p}^{\rm S}_i\vert^2.
\end{equation}
Inserting \eqn{eqn:H_har_2} into \eqn{eqn:equi_moment}, we obtain for the
momenta of the two particles:
\begin{equation}
    \Bigl\langle p_i^{\rm S}\frac{\partial \mathcal{H}_{\rm S}}
    {\partial p_i^{\rm S}} \Bigr\rangle
    = \Bigl\langle \frac{\bigl(p_i^{\rm S}\bigr)^2}{m_h} \Bigr\rangle
    = k_B T, \qquad i = 1,2,
\end{equation}
so that
\begin{equation}
    \bigl\langle \bigl(p_i^{\rm S}\bigr)^2 \bigr\rangle
    = m_h k_B T, \qquad i = 1,2.
\end{equation}
For $x = r_{12}$, the virial theorem in \eqn{eqn:equi_position} becomes
\begin{equation}
  \Bigl\langle r_{12}\,\frac{\partial \mathcal{H}_{\rm S}}{\partial r_{12}} \Bigr\rangle
  = 2a \bigl\langle r_{12}^2 \bigr\rangle
  = k_B T \biggl(1 +\frac{\int r_{12}\,\exp\bigl(-\beta a r_{12}^2\bigr)\,\frac{\partial \Phi(r_{12};T_{\rm R}, p_{\rm R})}{\partial r_{12}}\,dr_{12}}{\bar{Z}_{\rm S}}\biggr).
\end{equation}

\section{Temperature relaxation of Temperature under Langevin dynamics}\label{sec:langevin}
We derive \eqn{eqn:damping_eq} for the instantaneous temperature of a system obeying Langevin dynamics. In the derivation, the ensemble average $\langle\cdot\rangle$ denotes an average over many independent realizations of the same Langevin dynamics, i.e., repeated measurements performed at a fixed observation time $t$ but with different noise histories. Although temperature is strictly an equilibrium concept, during relaxation we introduce an effective (time-dependent) temperature through the instantaneous ensemble-averaged kinetic energy, $T(t) \equiv \frac{m}{k_B}\langle v^2(t)\rangle$.
This $T(t)$ should be interpreted as a kinetic-energy-based diagnostic rather than a statement of global thermodynamic equilibrium; it coincides with the bath temperature $T_{\mathrm{target}}$ only in the long-time limit.

We begin with the 1D Langevin equation for a particle of mass $m$,
\begin{equation}\label{eqn:langevin}
m\frac{dv}{dt} = -m\gamma v + R(t),
\end{equation}
where $\gamma$ is the damping coefficient and $R(t)$ is a zero-mean random force satisfying the fluctuation–dissipation relation \cite{tuckerman2010,tadmor2011}:
\begin{equation}
\langle R(t) \rangle = 0, \qquad \langle R(t) R(t') \rangle = 2m\gamma k_B T_{\mathrm{target}}\,\delta(t-t'),
\end{equation}
where $T_{\mathrm{target}}$ is the temperature of the heat bath. The solution of \eqn{eqn:langevin} is
\begin{equation}
v(t) = v(0)e^{-\gamma t} + \frac{1}{m}\int_0^t e^{-\gamma (t-s)} R(s)\,ds.
\end{equation}
The mean-square velocity $\langle v^2(t)\rangle$ follows as
\begin{align}
\langle v^2(t) \rangle
&=
\biggl\langle \bigl[v(0)e^{-\gamma t}
+ \frac{1}{m}\int_0^t e^{-\gamma (t-s)}R(s)\,ds \bigr]^2 \biggr\rangle \\
&=
\langle v(0)^2 \rangle e^{-2\gamma t}
+ \frac{1}{m^2} \int_0^t\!\!\int_0^t  e^{-\gamma (t-s)} e^{-\gamma (t-s')} \langle R(s)R(s')\rangle\,ds\,ds',
\end{align}
where the cross term vanishes since $\langle v(0) R(s) \rangle = v(0)\langle R(s) \rangle = 0$.
Using the noise correlation,
\begin{equation}
\langle R(s)R(s')\rangle
= 2m\gamma k_B T_{\mathrm{R}}\delta(s-s'),
\end{equation}
the double integral reduces to a single integral:
\begin{equation}
\langle v^2(t) \rangle
= \langle v(0)^2\rangle e^{-2\gamma t}
+ \frac{2\gamma k_B T_{\mathrm{target}}}{m}
\int_0^t e^{-2\gamma (t-s)} ds.
\end{equation}
The remaining integral yields
\begin{equation}
\int_0^t e^{-2\gamma (t-s)}ds
= \frac{1 - e^{-2\gamma t}}{2\gamma},
\end{equation}
so the full expression becomes
\begin{equation}
\langle v^2(t) \rangle = \langle v(0)^2\rangle e^{-2\gamma t} + \frac{k_B T_{\mathrm{target}}}{m} \left( 1 - e^{-2\gamma t} \right).
\end{equation}
In 1D case, the temperature is defined from the average
kinetic energy,
\begin{equation}
\frac{1}{2}m\langle v^2(t) \rangle = \frac{1}{2}k_B T(t),
\end{equation}
which gives
\begin{equation}
T(t) = \frac{m}{k_B}\,\langle v^2(t) \rangle.
\end{equation}
Thus the system temperature relaxes exponentially toward $T_{\mathrm{target}}$, as
\begin{equation}
T(t) = T_{\mathrm{target}} + \bigl(T_{\mathrm{init}} - T_{\mathrm{target}}\bigr) e^{-2\gamma t},
\end{equation}
where $T_{\mathrm{init}} = \frac{m}{k_B}\,\langle v^2(0) \rangle$ is the initial temperature of the system.

\printbibliography